\documentclass[twocolumn,showpacs,preprintnumbers,amsmath,amssymb]{revtex4}

\usepackage{epsfig}

\begin{document}

\preprint{APS/123-QED}

\title{
The helicity amplitudes A$_{1/2}$ and A$_{3/2}$ for the
D$_{13}(1520)$ resonance obtained from the $\vec{\gamma} \vec{p}
\to p \pi^0$ reaction}

%
%

\author{J.~Ahrens$^9$,
S.~Altieri$^{15,16}$, J.R.M.~Annand$^6$, G.~Anton$^3$,
H.-J.~Arends$^9$\footnote{corresponding author: e-mail address
arends@kph.uni-mainz.de}, K.~Aulenbacher$^9$, R.~Beck$^9$,
C.~Bradtke$^1$, A.~Braghieri$^{15}$, N.~Degrande$^{4}$,
N.~d'Hose$^{5}$, D.~Drechsel$^9$, H.~Dutz$^2$,  S.~Goertz$^1$,
P.~Grabmayr$^{17}$, K.~Hansen$^8$, J.~Harmsen$^1$,
D.~von~Harrach$^9$, S.~Hasegawa$^{13}$, T.~Hasegawa$^{11}$,
E.~Heid$^9$, K.~Helbing$^3$, H.~Holvoet$^4$, L.~Van~Hoorebeke$^4$,
N.~Horikawa$^{14}$, T.~Iwata$^{13}$, O.~Jahn$^9$,
P.~Jennewein$^9$, T.~Kageya$^{14}$,
S.~Kamalov$^{9}$\footnote{permanent address: Laboratory of
Theoretical Physics, JINR Dubna, Moscow, Russia}, B.~Kiel$^3$,
F.~Klein$^2$, R.~Kondratiev$^{12}$, K.~Kossert$^7$,
J.~Krimmer$^{17}$, M.~Lang$^9$, B.~Lannoy$^4$, R.~Leukel$^9$,
V.~Lisin$^{12}$, T.~Matsuda$^{11}$, J.C.~McGeorge$^6$,
A.~Meier$^1$, D.~Menze$^2$, W.~Meyer$^1$, T.~Michel$^3$,
J.~Naumann$^3$, A.~Panzeri$^{15,16}$, P.~Pedroni$^{15}$,
T.~Pinelli$^{15,16}$, I.~Preobrajenski$^{9,12}$, E.~Radtke$^1$,
E.~Reichert$^{10}$, G.~Reicherz$^1$, Ch.~Rohlof$^2$, G.
Rosner$^6$, D.~Ryckbosch$^4$, M.~Sauer$^{17}$, B.~Schoch$^2$,
M.~Schumacher$^7$, B.~Seitz$^7$\footnote{present address: II.
Physikalisches Institut, Universit\"at Gie{\ss}en},
T.~Speckner$^3$, N.~Takabayashi$^{13}$, G.~Tamas$^9$,
A.~Thomas$^9$, L.~Tiator$^9$, R.~van de Vyver$^4$,
A.~Wakai$^{14}$, W.~Weihofen$^7$, F.~Wissmann$^7$,
F.~Zapadtka$^7$, G.~Zeitler$^3$\\
(GDH- and A2- Collaborations)}
%
\affiliation{$^1$  Inst. f\"ur Experimentalphysik,
    Ruhr-Universit\"at Bochum,  D-44801 Bochum, Germany}
\affiliation{$^2$ Physikalisches Institut, Universit\"at Bonn,
    D-53115 Bonn, Germany} \affiliation{$^3$  Physikalisches Institut,
    Universit\"at Erlangen-N\"urnberg, D-91058 Erlangen, Germany}
\affiliation{$^4$ Subatomaire en Stralingsfysica, Universiteit
    Gent, B-9000 Gent, Belgium} \affiliation{$^5$  CEA Saclay,
    DSM/DAPNIA/SPhN, F-91191 Gif-sur-Yvette Cedex, France}
\affiliation{$^6$ Department of Physics \& Astronomy, University
    of Glasgow, Glasgow G12 8QQ, U.K.}
\affiliation{$^7$  II.Physikalisches Institut,
    Universit\"at G\"ottingen, D-37073 G\"ottingen, Germany}
\affiliation{$^8$ Department of Physics, University of Lund,
    Lund, Sweden}
\affiliation{$^9$  Institut f\"ur Kernphysik, Universit\"at Mainz,
    D-55099 Mainz, Germany}
\affiliation{$^{10}$  Institut f\"ur Physik, Universit\"at Mainz,
    D-55099 Mainz, Germany}
\affiliation{$^{11}$  Faculty of Engineering, Miyazaki University,
    Miyazaki, Japan}
\affiliation{$^{12}$  INR, Academy of Science, Moscow, Russia}
\affiliation{$^{13}$  Department of Physics, Nagoya University,
    Chikusa-ku, Nagoya, Japan}
\affiliation{$^{14}$  CIRSE, Nagoya University,  Chikusa-ku,
    Nagoya, Japan}
\affiliation{$^{15}$  INFN, Sezione di Pavia, I-27100 Pavia,
    Italy}
\affiliation{$^{16}$ Dipartimento di Fisica Nucleare e Teorica,
    Universit\`a di Pavia, I-27100 Pavia, Italy}
\affiliation{$^{17}$
    Physikalisches Institut, Universit\"at T\"ubingen, D-72076
    T\"ubingen, Germany}
%
%

\date{\today}

\begin{abstract}
The helicity dependence of the $\vec{\gamma} \vec{p} \to p \pi^0$
reaction has been measured for the first time in the photon
energy range from 550 to 790\,MeV. The experiment, performed at
the Mainz microtron MAMI, used a 4$\pi$-detector system, a
circularly polarized, tagged photon beam, and a longitudinally
polarized frozen-spin target. These data are predominantly
sensitive to the $D_{13}(1520)$ resonance and are used to
determine its parameters.
\end{abstract}
\pacs{PACS number(s): 13.60.Le, 14.20.Gk, 25.20.Lj }

\maketitle

{\it I. Introduction.} -- Over many years, measurements of the
accessible observables in $\eta$ and single-pion photoproduction
have been the basis of theoretical activity aiming to extract the
properties of the baryon resonances beyond the $\Delta$. For
example, the properties of the $S_{11}(1535)$ resonance, which
dominates $\eta$ photoproduction near threshold, can be extracted
from the measurements of total and differential cross sections
\cite{Krus2,Krus3} without strong model-dependence.

The situation is not so straightforward for the $D_{13}(1520)$
resonance, since in both single-pion and $\eta$ photoproduction
other resonances also contribute. In fact, the photo-decay
amplitudes $A_{1/2}$ and $A_{3/2}$ for this resonance, extracted
using the VPI~\cite{SAID} and Glasgow~\cite{Glas} partial wave
analyses of single-pion production, are significantly different
from those evaluated by Mukhopadhyay {\it et al.}~\cite{Mukh} and
Tiator {\it et al.}~\cite{Tiat} who used mainly $\eta$ production
data. This discrepancy can be resolved by using the selectivity of
polarization observables which allow the small, non-dominant
resonances to be discerned via their interference with the
dominant multipoles.

\begin{figure}[h]
\epsfig{file=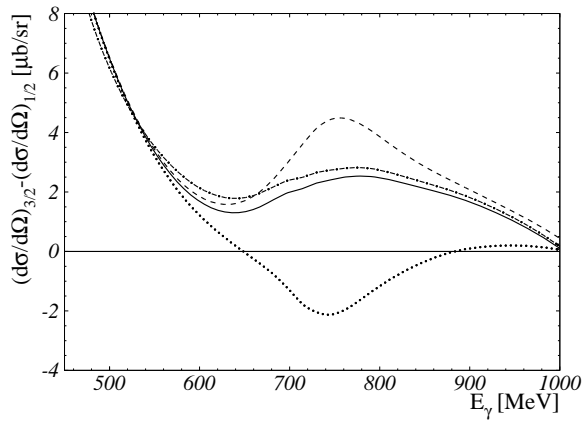,width=8cm} \caption{\label{fig:dx31_mot}
Energy dependence of the helicity dependent differential cross
section for $\vec{\gamma} \vec{p} \to p \pi^0$ at
$\theta^*=90^\circ$ as described by the UIM model~\cite{UIM}. The
curves represent: standard solution (full), no D$_{13}$(1520)
(dotted), no S$_{11}$(1535) (dashed), no P$_{11}$(1440)
(dash-dotted).}
\end{figure}

As is well-known, nine single and double polarization observables
have  to be measured to carry out a completely model-independent
multipole analysis of single-pion photoproduction. However, as
shown by Beck {\it et al.}~\cite{Beck}, some constraints can be
applied in order to perform an almost model-independent analysis
with fewer observables. A typical constraint is the low partial
waves approximation, which can be applied only in a limited energy
range. With increasing photon energy, the measurement of a more
comprehensive set of single and double polarization data becomes
very important. The sensitivity of an observable to small
multipoles can reflect a corresponding sensitivity to the more
weakly excited resonances. Figure \ref{fig:dx31_mot} illustrates
such sensitivity in the energy region from 450\,MeV up to 1\,GeV
for the  helicity dependent differential cross section
$(d\sigma/d\Omega)_{3/2}-(d\sigma/d\Omega)_{1/2}$ for the
$\vec{\gamma} \vec{p} \to p\pi^0$ channel. This was obtained
using circularly polarized photons and longitudinally polarized
nucleons and the subscripts $3/2$ and $1/2$ indicate the relative
nucleon-photon spin configurations, parallel and antiparallel,
respectively. This cross section difference is plotted as a
function of the photon energy at $\theta^*=90^\circ$, where
$\theta^*$ is the  pion angle in the CM-system. The full curve
represents the standard solution of the Unitary Isobar Model
(UIM)~\cite{UIM}, while the dotted, dashed and dashed-dotted
curves represent  solutions in which the coupling constant of the
$D_{13}(1520)$, the $S_{11}(1535)$ and the $P_{11}(1440)$
resonances, respectively, was set to zero. The difference between
the standard and modified solutions indicates the sensitivity of
this observable to the different resonances. As is clearly seen
in Figure~\ref{fig:dx31_mot}, the influence of the $D_{13}$
resonance is rather strong as the cross section difference even
changes sign. By contrast, the sensitivity to $P_{11}$ is almost
negligible and the sensitivity to $S_{11}$ is not very
pronounced. This observable is therefore well suited to extract
the parameters of the $D_{13}$ resonance.

In this letter, we present the first results for the helicity
dependent differential cross section of the $\vec{\gamma} \vec{p}
\to p \pi^0$ reaction in the energy range between 550 and
790\,MeV. These data were obtained during the GDH experiment
\cite{GDH1,GDH2} at the Mainz microtron MAMI, which studied the
helicity structure of the partial cross sections and their
contributions to the Gerasimov-Drell-Hearn sum rule.

{\it II. Experimental setup.} -- The main characteristics of the
experimental setup are summarized here, but more details may be
found in Refs. \cite{GDH1} and \cite{Mac}. The experiment was
conducted with the tagged photon facility~\cite{tag} at the MAMI
accelerator in Mainz. Circularly polarized photons were produced
by bremsstrahlung of longitudinally polarized
electrons~\cite{Aule}. The electron polarization (routinely about
75\%) was monitored during the data taking by a M\o ller
polarimeter.

Longitudinally polarized protons were provided by a frozen-spin
butanol ($C_4H_9OH$) target \cite{Brad}. The proton polarization
was measured using NMR techniques. Maximum polarization values
close to 90\% were obtained with a relaxation time of about
200\,h.

Photoemitted hadrons were registered by the large acceptance
detector DAPHNE~\cite{Audi}. DAPHNE essentially is a charged
particle tracking detector with cylindrical symmetry.  It covers
polar angles $\theta_{lab}=21^\circ$ to $159^\circ$.

{\it III. Data analysis.} -- The identification methods for
hadrons in DAPHNE have been described previously in detail
\cite{Mac,Brag} and only the main features will be recalled here.

Protons were identified using the range method \cite{Brag} making
simultaneous use of all of the charged particle energy losses in
the DAPHNE scintillator layers to discriminate between protons and
$\pi^\pm$ and to determine their kinetic energies. However, the
range method can be applied only to particles stopped inside
DAPHNE. This condition is satisfied by most of the protons
stemming from the $p\pi^0$ channel; only protons emitted with
$\theta_{lab} < 25^\circ$ and at $E_\gamma > 700$\,MeV cannot be
identified.

The presence of a single charged track recognized as a proton was
used as the signature for the $p\pi^0$ channel. The main
background in this case originates from the $p\pi^0\pi^0$ and
$p\pi^+\pi^-$ channels. The separation between the single and
double photoproduction channels was obtained from the analysis of
the missing mass spectrum $\gamma p \to p X$ \cite{2pi}.

The absolute efficiency of the $p\pi^0$ channel identification was
evaluated using a GEANT based simulation and found to be between
85\% and 95\%.

\begin{figure}[h]
\epsfig{file=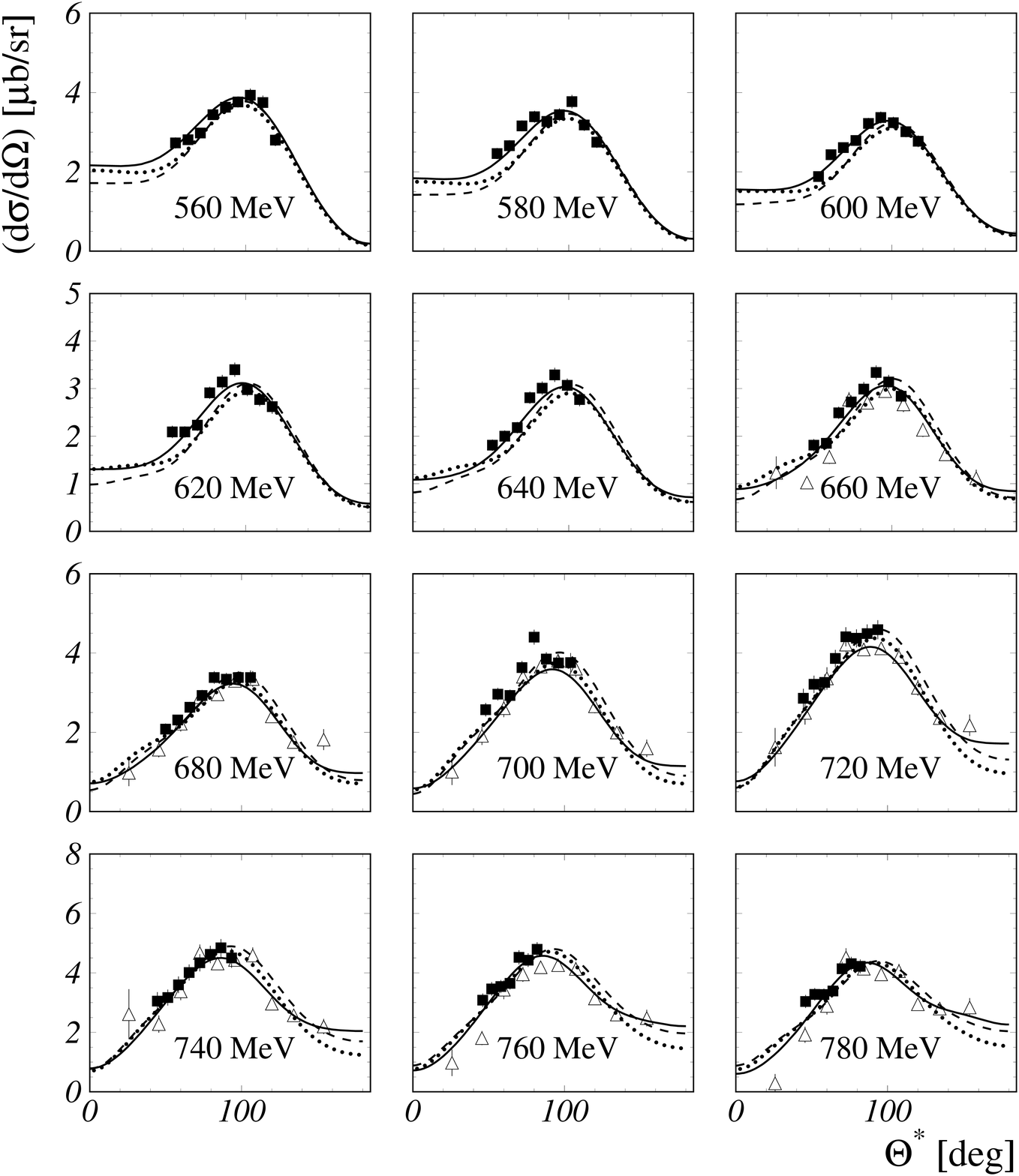,width=9.1cm}
\caption{\label{fig:dsg_data} The measured unpolarized
differential cross section for the $\gamma p \to p\pi^0$ reaction
for photon energies from 560 to 780 MeV (solid squares) is
compared to the measurement of Ref. \cite{Krus1} (open triangles)
and to the SAID~\cite{SAID} (solid curve) and the UIM~\cite{UIM}
(dashed curve) analyses. The dotted curve represents the modified
solution of UIM (see text). The errors shown are statistical
only.}
\end{figure}

Prior to the main experiment, data for detector calibration and
for tests of the analysis methods were taken with the same
apparatus using an unpolarized pure liquid hydrogen target. The
total unpolarized cross sections for $\gamma p \to n\pi^+$ and
$\gamma p \to p\pi^0$ in the $\Delta$ region were found to be in
a good agreement \cite{GDH1} with previously published data and
with predictions of multipole analyses. This confirms that the
detector response is well understood. Figure~\ref{fig:dsg_data}
shows the differential cross sections for $\gamma p \to p\pi^0$
in the energy range 550\,MeV $< E_\gamma <$ 790\,MeV \cite{Preo},
compared to the data of Ref. \cite{Krus1} and to the results of
the UIM~\cite{UIM} model and the SAID~\cite{SAID} multipole
analysis. The agreement shows that the detector response is
similarly well understood in this higher energy region.

As discussed previously \cite{GDH1}, in the analysis of data
taken using the butanol target, the background contribution of the
reactions on C and O nuclei could not be fully separated
event-by-event from the polarized H contribution. However, this
background from spinless nuclei is not polarization dependent and
cancels when the difference between events in the 3/2 and 1/2
states is taken. For this reason only the differential cross
section difference can be directly extracted from the measurement
with the butanol target.

\begin{figure}[h]
\epsfig{file=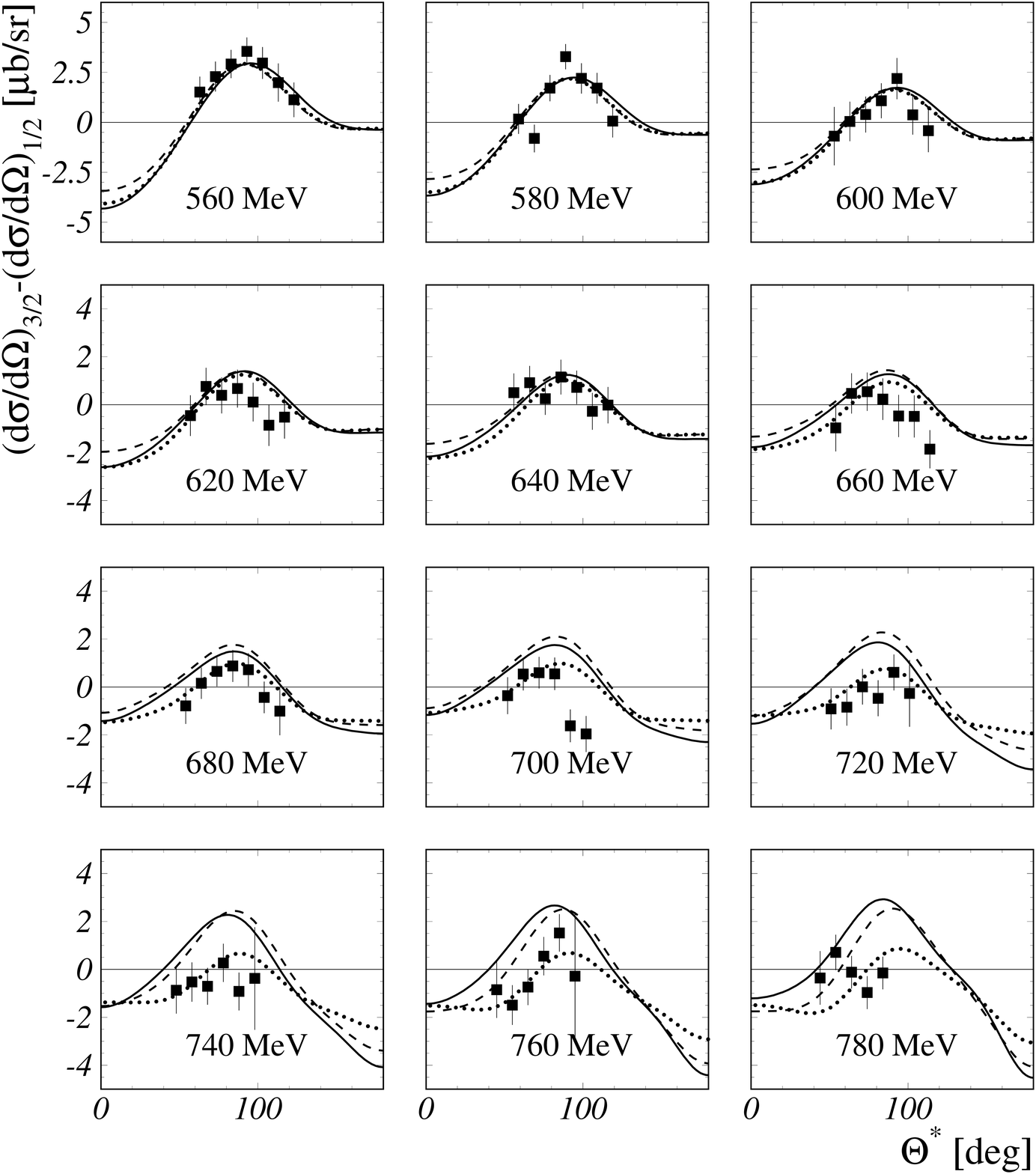,width=9.1cm}
\caption{\label{fig:dx31_data} The measured differential cross
section difference for $\vec{\gamma} \vec{p} \to p\pi^0$ (solid
squares). Curves as in Fig. \ref{fig:dsg_data}.}
\end{figure}

{\it IV. Results and discussion.} -- By using the methods
described above, the helicity dependent differential cross section
$(d\sigma/d\Omega)_{3/2}-(d\sigma/d\Omega)_{1/2}$ was obtained as
a function of pion angle $\theta^*$ in the CM-system in the
photon energy region from 550\,MeV up to 790\,MeV \cite{Preo}.
The results are presented in Figure~\ref{fig:dx31_data}. The
errors shown are statistical only. The systematic uncertainties
contain contributions from charged particle identification
(2.5\%), photon flux normalization (2\%), photon polarization
(3\%) and target polarization (1.6\%). The addition of these
errors in quadrature leads to a total systematic error of about
5\%.

At lower photon energies, the data are in a good agreement with
model predictions, but there is a clear systematic discrepancy
when the $D_{13}(1520)$ resonance is approached. In order to
extract information about this resonance, a fit of our
unpolarized and polarized differential cross sections, based on
the UIM \cite{UIM}, has been performed.

Since our data  cover only the angular region around $\theta^* =
90^\circ$, additional cross sections were included in the fit to
reduce the model dependence of this procedure. These were the
$p\pi^0$ data from \cite{Leuk}, which contain unpolarized
differential cross sections with full polar angle coverage in the
energy region 200\,MeV $< E_\gamma <$ 790\,MeV, the photon
asymmetry $\Sigma$ for 250\,MeV $< E_\gamma <$ 440\,MeV, and the
$n\pi^+$ and $p\pi^0$ polarized differential cross sections in the
$\Delta$ resonance region (180\,MeV  $< E_\gamma <$ 450\,MeV),
measured in the GDH experiment \cite{Preo}. In this way, all the
resonant partial wave amplitudes up to a photon energy of 800 MeV
could be simultaneously determined in the fit.

Within the UIM framework, seven free parameters have been used in
our fit: six resonance couplings and the
pseudoscalar-pseudovector mixing parameter (PS/PV), which mostly
affects the amplitudes $E_{0+}^{1/2}$ and $M_{1-}^{1/2}$, see Ref.
\cite{UIM}. A modification of the resonance couplings only
affects the imaginary part of the resonance amplitude in the
corresponding partial wave. The resulting modification factors
for the $D_{13}$ resonance compared to the standard UIM couplings
have been found to be  1.11$\pm$0.01 ($M_{2-}^{1/2}$) and
0.81$\pm$0.01 ($E_{2-}^{1/2}$) while the other parameters
remained unchanged within the fitting errors. This modified UIM
solution is shown in Figures \ref{fig:dsg_data} and
\ref{fig:dx31_data} by the dotted curves.

The $D_{13}(1520) \to p\gamma$ helicity amplitudes $A_{1/2}$ and
$A_{3/2}$ were then evaluated from the modified UIM solution. In
principle, the transition from the electric and magnetic
representation to helicity amplitudes is model dependent since a
separation between resonant and background multipole components
has to be performed. Once the separation is done, the standard
recipe described in Ref. \cite{arnd} can be used. Since the
$D_{13}$ partial wave amplitudes are almost purely imaginary at
the resonance position, the model dependence is weak. Because the
background is real, only the imaginary partial wave amplitudes
are required to calculate the helicity couplings from electric and
magnetic partial waves. This situation is related to the nearly
perfect Breit-Wigner shape of the $D_{13}(1520)$. Taking a
resonance position of 1520\,MeV, a resonance width of 120\,MeV,
and a pion branching ratio of 0.55 (from PDG \cite{PDG}), the
helicity amplitudes $A_{1/2}$ and $A_{3/2}$ were found to be
-0.038$\pm$ 0.003\,GeV$^{-1/2}$ and 0.147 $\pm$
0.010\,GeV$^{-1/2}$, respectively. The errors are a combination
of the statistical fitting errors and the estimated model errors
due to the uncertainties in the $D_{13}$ resonance parameters.
Using the same method, the helicity amplitudes were extracted
from the standard UIM (MAID2000) solution and from the SAID (SM01)
solution. These results are summarized in Table~\ref{Table}
together with the latest PDG estimate \cite{PDG}.

\begin{table}[h]
\begin{center}
\begin{tabular}{|c|c|c|}
\hline
\hbox to 5mm {\hfill} Solution \hbox to 5mm {\hfill} &
          \hbox to 7mm {\hfill}  $A_{1/2}$ \hbox to 7mm {\hfill} &
      \hbox to 7mm {\hfill}  $A_{3/2}$ \hbox to 7mm {\hfill} \\
\hline
\hline
standard UIM & -0.017 & 0.164 \\
SAID  & -0.016 & 0.167 \\
PDG estimate & -0.024 $\pm$ 0.009 & 0.166 $\pm$ 0.005 \\
\hline
modified UIM & -0.038 $\pm$ 0.003 & 0.147 $\pm$ 0.010\\
\hline
\end{tabular}
\caption{The $D_{13}$ helicity amplitudes $A_{1/2}$, $A_{3/2}$
for the proton (in units of GeV$^{-1/2}$) estimated from the
modified UIM analysis, are compared to the standard UIM
(MAID2000) solution, SAID (SM01) analysis and PDG, see text.}
\label{Table}
\end{center}
\end{table}

\begin{figure}[ht]
\epsfig{file=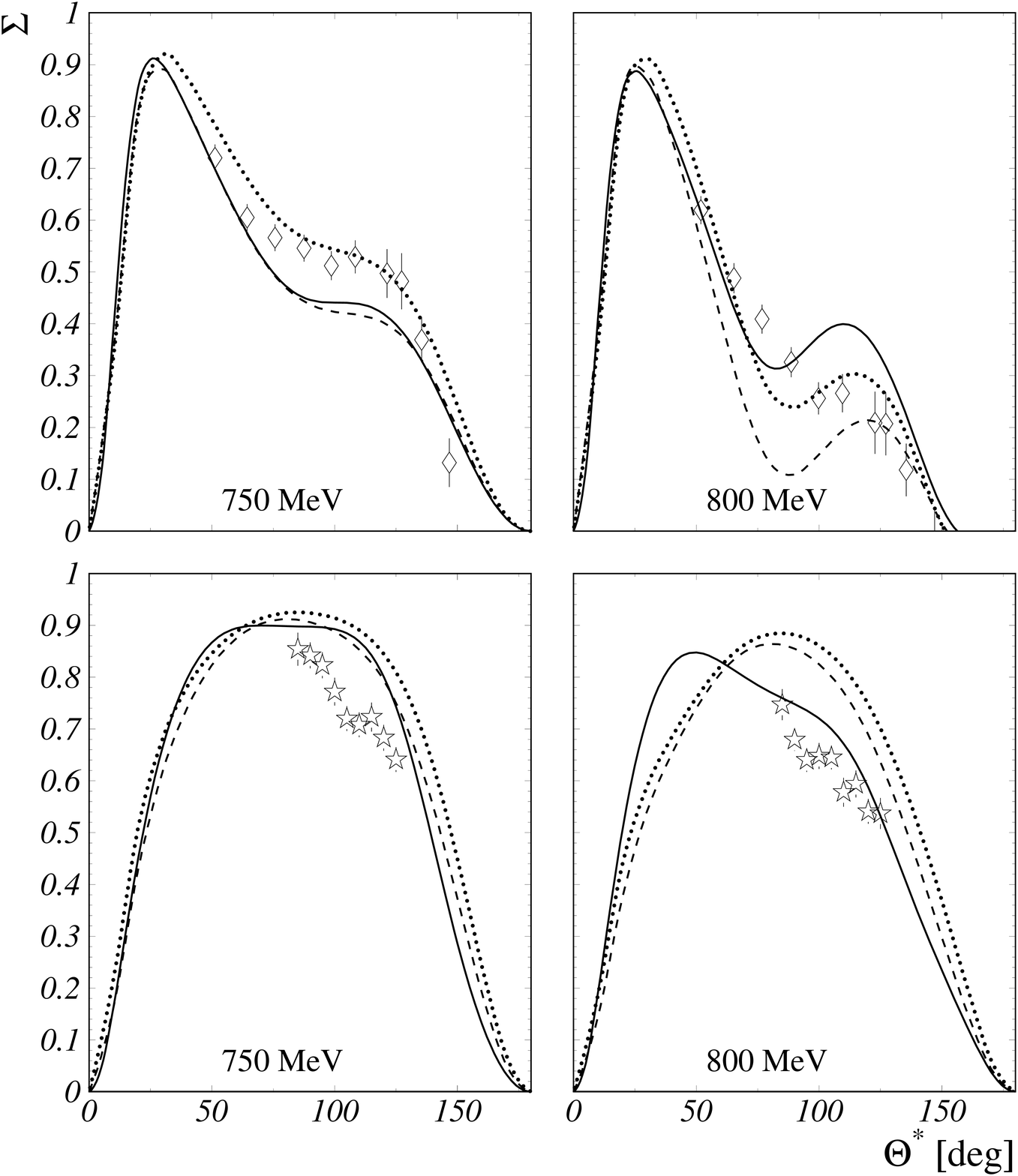,width=9cm}
 \caption{\label{fig:s_data} Photon asymmetry ($\Sigma$)
 measurements at E$_\gamma$=750 and 800 MeV
for  $\gamma p \to n\pi^+$ \cite{Ajak} (top) and for $\gamma p
\to p\pi^0$ \cite{Adam} (bottom). Curves as in Fig.
\ref{fig:dsg_data}.}
\end{figure}

As pointed out by Workman {\it et al.} \cite{Work}, the photon
asymmetry $\Sigma$ is also quite sensitive to the parameters of
the $D_{13}(1520)$ resonance. This observable has been recently
measured for the $\gamma p \to p\pi^0$ channel in Yerevan
\cite{Adam} and for the $\gamma p \to n\pi^+$ channel at GRAAL
\cite{Ajak}. A comparison has therefore been made between our
modified UIM solution and these new data, as shown in
Figure~\ref{fig:s_data}. In the same figure, the results of the
standard UIM and SAID solutions are also plotted. The modified
UIM solution is in satisfactory agreement with the $n\pi^+$ data,
but tends to disagree with the $p\pi^0$ data. However,
preliminary data from GRAAL for this latter channel~\cite{bart}
agree better with the theories and in this case the sensitivity
to the parameters of the $D_{13}$ resonance is small.

In conclusion, our first data on the helicity dependent cross
sections for $\pi^0$ photoproduction can be used to determine the
photo coupling parameters of the $D_{13}$ resonance, due to the
almost exclusive sensitivity of the helicity difference to this
resonance.  However, other data are needed to reconstruct a
sufficient number of partial waves. Existing unpolarized cross
sections and photon asymmetry data for the $\gamma p \to n\pi^+$
channel seem to be good candidates for this purpose. Our data
imply that $A_{1/2}$ is larger (in absolute value) by about 60\%
and $A_{3/2}$  is smaller by about 12\% compared to the standard
PDG values.

The authors wish to acknowledge the excellent support of the
accelerator group of MAMI. This work was supported by the
Deutsche Forschungsgemeinschaft (SFB 201, SFB 443 and
Schwerpunktprogramm 1034), the INFN--Italy,  the FWO
Vlaanderen--Belgium, the IWT--Belgium, the UK Engineering and
Physical Science Research Council, the DAAD, and the
Grant-in-Aid, Monbusho, Japan.


\begin{thebibliography}{99}

\bibitem{Krus2}
B. Krusche {\it et al.}, Phys. Lett.  {\bf B397},
171 (1997).

\bibitem{Krus3}
B. Krusche {\it et al.}, Phys. Rev. Lett.  {\bf 74},
3736 (1995).

\bibitem{SAID}
R.A. Arndt {\it et al.}, Phys. Rev. C {\bf 53}, 430 (1996), SAID
(GWU) solution SM01, R.A. Arndt, I. Strakovsky and R. Workman (to
be published).

\bibitem{Glas}
R.L. Crawford, W.T. Morton, Nucl. Phys. B {\bf 211},
1, (1983).

\bibitem{Mukh}
N.C. Mukhopadhyay {\it et al.}, Phys. Lett. B {\bf 444},
7 (1998).

\bibitem{Tiat}
L. Tiator {\it et al.}, Phys. Rev. C {\bf 60},
035210 (1999)

\bibitem{Beck}
R. Beck {\it et al.}, Phys. Rev. C {\bf 61}, 035204 (2000), R.
Beck {\it et al.},  Phys. Rev. Lett. {\bf 78}, 606 (1997).

\bibitem{UIM}
D. Drechsel {\it et al.}, Nucl. Phys.  {\bf A645},
145 (1999).

\bibitem{GDH1}
J. Ahrens {\it et al.}, Phys. Rev. Lett.  {\bf 84}, 5950 (2000).

\bibitem{GDH2}
J. Ahrens {\it et al.}, Phys. Rev. Lett.  {\bf 87},
022003 (2001).

\bibitem{Mac}
M. MacCormick {\it et al.}, Phys. Rev. C {\bf 53},
41 (1996).

\bibitem{tag}
I. Anthony {\it et al.}, Nucl. Instrum. Methods Phys. Res., Sect. A  {\bf 301},
230 (1991); S.J. Hall {\it et al.}, Nucl. Instrum. Methods Phys. Res.,
Sect. A  {\bf 368}, 698 (1996).

\bibitem{Aule}
K. Aulenbacher {\it et al.}, Nucl. Instrum. Methods Phys. Res.,
Sect. A  {\bf 391}, 498 (1997).

\bibitem{Brad}
C. Bradtke {\it et al.}, Nucl. Instrum. Methods Phys. Res.,
Sect. A  {\bf 436}, 430 (1999), C. Bradtke, Ph.D. thesis, University of
Bonn, 2000.

\bibitem{Audi}
G. Audit {\it et al.}, Nucl. Instrum. Methods Phys. Res.,
Sect. A  {\bf 301}, 473 (1991).

\bibitem{Brag}
A. Braghieri {\it et al.}, Nucl. Instrum. Methods Phys. Res.,
Sect. A  {\bf 343}, 623 (1994).

\bibitem{2pi}
A. Braghieri {\it et al.} Phys. Lett. B {\bf 363},
46, (1995).

\bibitem{Preo}
I. Preobrajenski, Ph.D. thesis, University of Mainz, 2001, to be published.

\bibitem{Krus1}
B. Krusche {\it et al.}, Eur. Phys. J. A {\bf 6},
309 (1999).

\bibitem{Leuk}
R. Leukel, Ph.D Thesis, University of Mainz, 2001, to be published.

\bibitem{Work}
R. Workman {\it et al.}, Phys. Rev. C {\bf 62}, 048201 (2000).

\bibitem{arnd}
D. Arndt {\it et al.}, Phys. Rev. C {\bf 42},
1864 (1990).

\bibitem{PDG}
D.E. Groom {\it et al.} Eur. Phys. Jour. C {\bf 15},
1, (2001).

\bibitem{Ajak}
J. Ajaka {\it et al.}, Phys. Lett. B {\bf 475}, 372 (2000).

\bibitem{Adam}
F.V. Adamian {\it et al.}, Phys. Rev. C {\bf 63}, 054606 (2001).

\bibitem{bart}
O. Bartalini {\it et al.}, Prog. Part. Nucl. Phys. {\bf 44}, 423
(2000).


\end{thebibliography}
\end{document}